# A Simulation Experiment on a Built-In Self Test Equipped with Pseudorandom Test Pattern Generator and Multi-Input Shift Register (MISR)

Afaq Ahmad

Department of Electrical and Computer Engineering
College of Engineering, Sultan Qaboos University
P. O. Box 33, Postal Code 123; Muscat, Sultanate of Oman
Telephone: (+ 968) 2414 1327
Fax: (+968) 2441 3416
afaq@squ.edu.om

## ABSTRACT

*This paper investigates the impact of the changes of the characteristic polynomials and initial loadings, on behaviour of aliasing errors of parallel signature analyzer (Multi-Input Shift Register), used in an LFSR based digital circuit testing technique. The investigation is carried-out through an extensive simulation study of the effectiveness of the LFSR based digital circuit testing technique. The results of the study show that when the identical characteristic polynomials of order n are used in both pseudo-random test-pattern generator, as well as in Multi-Input Shift Register (MISR) signature analyzer (parallel type) then the probability of aliasing errors remains unchanged due to the changes in the initial loadings of the pseudo-random test-pattern generator.*

## KEYWORDS

*LFSR, MISR, BIST, Characteristic Polynomial, Primitive*

## 1.0 INTRODUCTION

Reliability is one of the main considerations in any circuit design. It involves a correct and predictable behaviour of the circuit according to design specifications over a sufficiently long period of time. To achieve this goal, the logic-circuit design is aimed at an error-free circuit operation. Unfortunately, in spite of all possible care being bestowed on design and simulation, hardware faults resulting from physical defects (e.g. mask defects, manufacturing process flaws) will occur in the hardware implementation of the logic circuit. Hence, when a fault occurs anyway, one must be able to detect the presence of the fault and, if desired to pinpoint its location. This task is accomplished by testing the circuit. System maintenance draws heavily upon the testing capability of the logic system [1], [2].

Digital circuit manufacturers are well aware of the need to incorporate testability features early in the design stage, or otherwise they have to incur higher testing costs, subsequently. An empirical relationship, that have been used for estimating the cost of finding a faulty chip, indicates that the cost increases by a factor of 10, as fault finding advances from one level to the next [3] – [7]. However, recent studies have shown that the cost of testing and fault finding, at system and field level, is higher than this factor of 10 and increases exponentially [7] – [9]. Thus, if a fault can be detected at chip or board level, then significantly larger costs per fault can be avoided. This is the prime reason that attention has now focused on providing testability at chip, module or even at board level.

Any test methodology usually consists of
- (i) A test strategy for generating the test-stimuli,
- (ii) A strategy for evaluating output responses, and
- (iii) Implementation mechanisms to realize the appropriate strategies in test-generation and response evaluation.

Present day philosophy to achieve economical and cost effective testing of Very Large Scale Integration (VLSI) components and systems is to provide on-chip testing. Though these techniques involve additional chip area for the added test circuitry, they have provided reasonable testability levels. In fact it has been reported that, for about 20% additional silicon area required, more than 98% of the chip design can be checked using structured Design For Testability (DFT) approaches [7] – [11].

As a natural outcome of the structured design approach for DFT, built-in testing has drawn considerable attention. Usually a built-in test methodology is defined as the one that incorporates both test-pattern generation and response data compression mechanisms internally in the chip itself. If this methodology is self-sufficient in detecting the faults of its internal test circuitry also, then such a methodology is referred to as Built-In Self-Test (BIST) in test literature. The main emphasis in BIST designs is that to provide close to one hundred percent testing of combinational circuits [10], [12]. In particular, pseudo-random test-pattern generation followed by the compression of response data by signature analysis has become a standard form of testing in BIST environment. Linear Feedback Shift Registers (LFSRs) have been proposed as an integral part of a sequential logic design, such that they can be used to both generate and compact the results of a test. Undoubtedly, an LFSR based pseudo-random test-pattern generation is an extremely simple tool for generating desired sequence as well as the length of the test-stimuli. Many estimation schemes are readily available for computing the length of test patterns where the desired sequence of the test-patterns can be obtained by the predetermined seed of the LFSRs. Further, the effective testing of large circuits uses the concept of 'pseudo-exhaustive testing' where the principle of divide and conquer is applied [13] – [19].

Difficulty arises when the resulting response data obtained from the Circuit Under Test (CUT) is compressed into small signatures using Signature Analyzer (SAZ). Although, this scheme of SAZ is easily implemented by an LFSR either in the form of Single Input Shift Register (SISR) – serial SAZ or MISR- parallel SAZ . But this leads to loss of information, due to the erroneous response patterns that get compressed into the same signature as the fault-free signature of the CUT. Thus, some of the faults might go undetected due to this masking phenomenon. This compression can further reduce the fault-coverage in the BIST scheme. This problem of error masking is called aliasing [20] – [25].

Methods to determine the extent of fault escape caused by a response compressor are not readily available. However, various attempts have been made to analyze and improve the basic SAZ's realization methods [26] – [33]. The end goal of the above schemes, individually, or with a combination of these, is to reduce the deception volume [26]. Methods that consider both, the test pattern generator and response data compressor factors in totality and simultaneously, in analyzing the aliasing behaviour of the circuit, are not available. There is a growing need for such an approach, which comprehensively looks at aliasing problems with respect to both test pattern generator and response data compressor and reflects the true aliasing characteristics. This is the prime justification of the research work in this area. Therefore, towards this direction a research work have been done [34], [35]. Through these papers, different studies were carried out to investigate the roles of characteristic polynomials used in Pseudo-Random Test-Pattern Generator (PRTPG) as well as in SAZ, and initial loading of PRTPG with the behaviour of aliasing errors of SAZ. The work contained in the papers [34], [35] used separate different structures (internal and external exclusive-OR types) of LFSRs. However, both the papers considered SISR type of SAZ. This paper is an extension of the previous research where MISR type of SAZ is used. In this work external exclusive-OR type and internal exclusive-OR type LFSR is used in PRTPG and MISR respectively. The results of the study show that the probability of aliasing errors remains unchanged due to the change in the initial loading when the identical characteristic polynomials of order n are used in both PRTPG, as well as in SAZ (MISR type).

## 2.0 MATHEMATICAL CHARACTERIZATION OF LFSR

In this section we consider briefly the mathematical background, definitions and theorems related to an LFSR. Details can be found in literatures [36] – [39]. Basic definitions and theorems are given below for the sake of completeness.

Let [A] represent the state transition matrix of order n × n, for an n stage LFSR shown in Figure 1. Assume the state at any time 't' be represented by vector $[Q(t)] = [q_n(t), ... ,q_j(t), ... ,q_2(t), q_1(t)]$ (which is effectively the contents of the LFSR) where each $q_j$ represents the state of the $j^{th}$ stage of the LFSR. Further, let the LFSR feedback stages be numbered from $C_0$ to $C_n$, proceeding in the same direction as the shifting occurs i.e. left to right. Let the present state of the LFSR be represented by [Q(t)] and, one clock later, the next state by [Q(t+1)]; then the relationship between the two states is given by Equation (1).

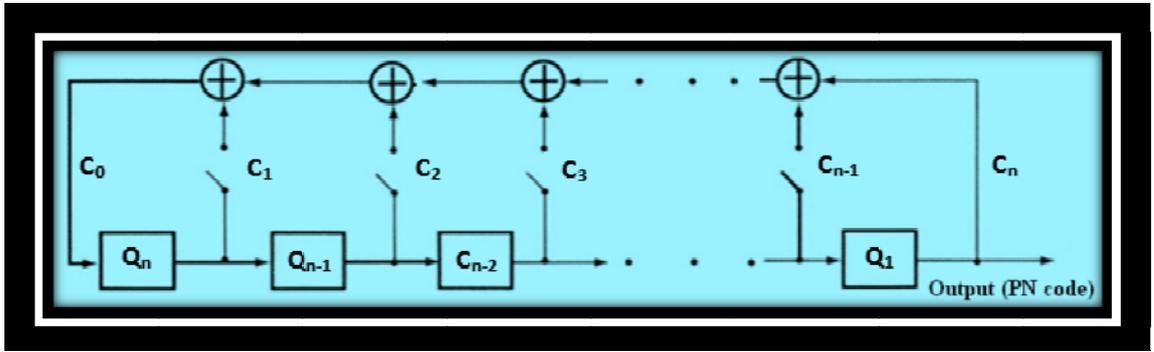

Figure 1. An n-bit LFSR model

$$\begin{bmatrix} q_n(t+1) \\ q_{n-1}(t+1) \\ q_{n-2}(t+1) \\ \vdots \\ q(t+1) \\ q(t+1) \end{bmatrix} = \begin{bmatrix} c_n & c_{n-1} & ... & c_2 & c_1 \\ 1 & 0 & ... & 0 & 0 \\ 0 & 1 & ... & 0 & 0 \\ \vdots & \vdots & \vdots & \vdots & \vdots \\ 0 & 0 & ... & 0 & 1 \\ 0 & 0 & ... & 1 & 0 \end{bmatrix} \begin{bmatrix} q_n(t) \\ q_{n-1}(t) \\ q_{n-2}(t) \\ \vdots \\ q(t) \\ q(t) \end{bmatrix}$$
(1)

where, $c_j$ = 0 or 1, for 1≤ j ≤ n-1 and $c_j$ = 1, for j = n. (2)

In Equation (2), the values of $C_j$ show the existence or absence of a feedback connection from the $j^{th}$ stage of the LFSR. Equation (1) can be written as

[Q(t+1)] = [A][Q(t)]   (3)

If [Q] = [Q(0)] represents a particular initial loading of the LFSR, then the sequence of states through which the LFSR will pass during successive times is given by

$[Q(t)], [A][Q(t)], [A]^2[Q(t)], [A]^3[Q(t)], …$

Let the matrix 'period' be the smallest integer p for which $[A]^p = I$, where I is an identity matrix. Then $[A]^p[Q(t)] = [Q(t)]$ for any non zero initial vector [Q(0)], indicating the 'cycle length (or period)' of the LFSR is p.
Thus, on the basis of this property of periodicity of LFSR and Equation (3), it follows that

$$[Q(t)] = [Q(t+p)] = [A]^p[Q(t)] \qquad (4)$$

**Definition 1**:
The cycle length p, for vector [Q(0)] = 0 is always 1, which is independent of matrix [A].

**Definition 2**: The period p of an n bit LFSR will only be maximal when $p = m = 2^n - 1$.

For the matrix [A] of the LFSR, the characteristic equation is given by
Determinant $[A - \lambda I] = 0$. Thus,

$$F(\lambda) = \begin{bmatrix} C_n - \lambda & C_{n-1} - \lambda & C_{n-2} - \lambda & ..C_{n-i} - \lambda & C_2 - \lambda & C_1 - \lambda \\ 1 & -\lambda & 0 & ..0 & 0 & 0 \\ \vdots & \vdots & \vdots & ...-\lambda & \vdots & \vdots \\ 0 & 0 & 0 & \cdots 1 & -\lambda & 0 \\ 0 & 0 & 0 & \cdots 0 & 1 & -\lambda \end{bmatrix} \qquad (5)$$

**Definition 3**: For the matrix [A] of an LFSR, the polynomial of {determinant $[A-\lambda I]$} is called the characteristic polynomial $F(\lambda)$ of the LFSR and can be written as

$$F(\lambda) = 1 + \sum_{j=1}^{n} C_j \lambda^j \qquad ; C_n = 1. \qquad (6)$$

Let, $T(\lambda)$ denote the characteristic polynomial of an n stage LFSR used in PRTPG. Let, {$a_{-1}, a_{-2},...,a_{-n+1}, a_{-n}$} be the initial state of the shift register. The sequence of numbers $a_0, a_1, a_2, ... a_q ...$ can be associated with a polynomial, called generating function $M(\lambda)$, by the rule

$$M(\lambda) = a_0 + a_1 \lambda + ... + a_q \lambda^q + ...$$

Let $\{a_q\} = a_0, a_1, a_2,...$ represent the output sequence generated by an LFSR used as PRTPG, where $a_i = 0$ or 1. Then this sequence can be represented as

$$M(\lambda) = \sum_{q=0}^{\infty} a_q \lambda^q \qquad (7)$$

From the structure of the type of the LFSR shown in Figure 2, it can be seen that if the current state of the $j^{th}$ flip-flop is $a_{q-j}$, for j = 1, 2, ... , n , then by the recurrence relation

$$a_q = \sum_{j=1}^{n} C_j a_{q-j} \qquad (8)$$

Substituting (8) in (7)

$$M(\lambda) = \sum_{q=0}^{\infty} \sum_{j=1}^{n} C_j a_{q-j} \lambda^q \qquad (9)$$

Or, by solving for generating function, it can be shown that M(λ) is given by

$$M(\lambda) = \frac{\sum_{j=1}^{n} C^{j} \lambda^{j} (a_{-j} \lambda^{-j} + a_{-j+1} \lambda^{-j+1} + \ldots + a_{-1} \lambda^{-1})}{1 + \sum_{j=1}^{n} C_{j} \lambda^{j}} \qquad (10)$$

$$M(\lambda) = \frac{\sum_{j=1}^{n} C^{j} (a_{-j} + a_{-j+1} \lambda + \ldots + a_{-1} \lambda^{j-1})}{1 + \sum_{j=1}^{n} C_{j} \lambda^{j}} \qquad (11)$$

Or,

$$M(\lambda) = \frac{B(\lambda)}{T(\lambda)} \qquad (12)$$

Thus, the PRTP (represented by polynomial M(λ), generated by the LFSR can be obtained through the long division of the function B(λ) by T(λ). Therefore, it can be implied from the Equation (11) that the generated sequence M(λ) is function of initial loading as well characteristic polynomial of the LFSR used in the realization of the PRTPG.

## 3.0 MULTI-INPUT SHIFT REGISTER (MISR) MODEL

Figure 2 shows a typical MISR configuration. This configuration of MISR is based on the internal-EXOR. In the figure, n denotes the length of the MISR, i.e., the number of FFs in the register. Also, the LFSR feedback stages be numbered from $C_n$ to $C_1$, are the binary coefficients of the characteristic polynomial P (of the MISR.

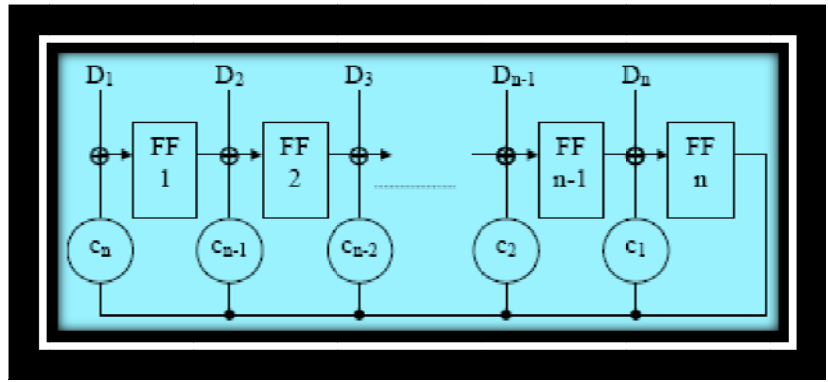

Figure 2. An n-bit MISR model

$$P(\lambda) = 1 + \sum_{j=1}^{n} C_{j} \lambda^{j} \qquad (13)$$

Let $f_i$ be the content of the i-th FF, then the state of the MISR

(i.e., the sequence $r = r_n + r_{n-1} + \ldots + r_2 + r_1$) can be represented by the state polynomial

$$s(\lambda) = r_n \lambda^{n-1} + r_{n-1} \lambda^{n-2} + \ldots + r_2 \lambda + r_1 \qquad (14)$$

Similarly, an n-bit input sequence ($d = D_n + D_{n-1} + \ldots + D_2 + D_1$) can be represented by the input polynomial.

$$d(\lambda) = D_n \lambda^{n-1} + D_{n-1} \lambda^{n-2} + \ldots + D_2 \lambda + D_1 \qquad (15)$$

The signature obtained by SAZ (any SISR or MISR) is defined as the final state of the register after the input sequence d has been entered into the register. Then the MISR signature can be given as

$$s(x) = D(x) \bmod P(x) \qquad (16)$$

Thus, the theory behind the use of the LFSR for SAZ is based on the concept of polynomial division process, where the remainder left in the register after the last bit of input data is sampled, corresponds to the final signature. Whereas, the output sequence from the $n^{th}$ bit of the LFSR defines the quotient, QO of the division. In general, the shift register is initialized by the reset or by parallel load function of the register, at a time of fault-free evaluation as well as every time when a new fault is injected in the CUT. Assume that the CUT is of combinational nature with n primary inputs and k primary outputs. If the initial state of the LFSR is all 0's, let the final state of the LFSR be represented by the polynomial $s(\lambda)$. Then it can be shown that these polynomials are related by the equation

$$\frac{d(\lambda)}{P(\lambda)} = QO(\lambda) + \frac{s(\lambda)}{P(\lambda)} \qquad (17)$$

where, $P(\lambda)$ is the characteristic polynomial of the LFSR structure used in MISR-SAZ. Hence an LFSR carries out polynomial division on the input data stream by the characteristic polynomial, producing an output stream corresponding to the quotient $QO(\lambda)$ and remainder $s(\lambda)$.

## 3.0 SIMULATION STUDY

The testing model employed in the simulation study for PRTPG and MISR is as described in section 2 and shown in Figures 1 and 2. Various combinational circuits have been simulated using the manufacturer's logical diagrams with gate level description. A single stuck-at fault model is assumed. Where, s-a-0 and s-a-1 faults are postulated on each individual, $N_L$, branches of the each CUT. In the case of each n-input CUT, identical all possible characteristic polynomials of order n are individually applied to PRTPG and MISR-SAZ as well. All possible initial loading of PRTPG, $2^n-1$, are exhausted to monitor the aliasing error behaviour of MISR-SAZ. These characteristic polynomials are generated using the algorithms developed and reported in papers [2], [40] and [41]. To make it more readable the simulation procedure used to study the aliasing behaviour is described below.

## Simulation procedure

```
Begin
(For an n-input CUT)
    NL = total number of branches in the CUT;
    NP = total number of possible characteristic polynomial of order n, over GF(2);
    L_i = is the periodicity of i^th characteristic polynomial of order n;
    NS = total number of possible initial loading {NS = 2^n-1, excluding S =[000..0]};
    RC = is the aliasing count;
 For i = 1 to NP, do
 For r = 1 to NS, do
 Begin
    Choose i^th characteristic polynomial for PRTPG as well as for SAZ {i.e.
    T_i(λ) and P_i(λ)}respectively;
    Choose r^th initial loading S_r;
    Generate test stimuli of length L_i;
    Choose circuit response d_g {fault-free circuit response};
    Compute s_g {fault-free signature};
  For t = 1 to 2NL, do
  Begin
    Initialize aliasing count RC;
    Choose circuit response d_ft {d_f1 has fault number 1 inserted, d_f2 has fault
    number 2 inserted, etc.};
    Compute signature s_ft {signature when the t^th fault is inserted};
    Compare s_ft, with s_g if s_ft ≠ s_g then, increment aliasing count RC;
  End do;
    Write aliasing count {one each for s_r, T_i (λ), P_i(λ)};
 End do;
End.
```

The above procedure is used to observe the effect of the characteristic polynomials used in PRTPG as well as in MISR-SAZ along with the initial loading of PRTPG on aliasing counts of MISR-SAZ. The aliasing counts for the circuits of Table 1 were monitored for the sets of all NP and NS of order n.

Table 1. Summary of simulated circuits

| Circuit Number | Module of IC Number | Circuit Specifications (n-inputs, k-outputs) | Number of Faults Injected | NP / NS of order n |
|---|---|---|---|---|
| C-1 | SN-74LS139 DUAL 2-TO-4 LINE DECODER/ DEMULTIPLEXER | 3-inputs; 4-outputs 9-gates | 58 | 2 / 7 |
| C-2 | SN-74LS82 2-BIT BINARY FULL ADDER | 5-inputs; 3-outputs 21-gates | 148 | 6 / 31 |
| C-3 | SN-74H87 4-BIT TRUE/ COMPLEMENT, ZEO/ ONE ELEMENT | 6-inputs; 4-outputs 14-gates | 64 | 6 / 63 |

The observed results demonstrates that when the identical characteristic polynomials are used in both the PRTPG and MISR-SAZ, then any change in initial loading of PRTPG does not change the value of aliasing count. Due the complexity of the result sets and space only a candidate of result for circuits summarized in Table 1 are shown in Tables 2 to 4. In Tables 2 – 4 the aliasing count is shown. These values of aliasing counts remain unchanged for all the possible changes of initial loading of PRTPG.

Table 2. Aliasing errors for all possible NS for circuit C-1

| PRTPG $T(\lambda)$ | MISR-SAZ $P(\lambda)$ | |
|---|---|---|
| | $1+\lambda+\lambda^4$ | $1+\lambda^3+\lambda^4$ |
| $1+\lambda+\lambda^3$ | 9 | 13 |
| $1+\lambda^2+\lambda^3$ | 13 | 9 |

Table 3. Aliasing errors for all possible NS for circuit C-2

| PRTPG $T(\lambda)$ | MISR-SAZ $P(\lambda)$ | |
|---|---|---|
| | $1+\lambda+\lambda^3$ | $1+\lambda^2+\lambda^3$ |
| $1+\lambda^3+\lambda^5$ | 4 | 12 |
| $1+\lambda^2+\lambda^5$ | 17 | 15 |
| $1+\lambda^2+\lambda^3+\lambda^4+\lambda^5$ | 13 | 9 |
| $1+\lambda+\lambda^3+\lambda^4+\lambda^5$ | 14 | 16 |
| $1+\lambda+\lambda^2+\lambda^4+\lambda^5$ | 10 | 18 |
| $1+\lambda+\lambda^2+\lambda^3+\lambda^5$ | 7 | 11 |

Table 4. Aliasing errors for all possible NS for circuit C-3

| PRTPG T($\lambda$) | MISR-SAZ P($\lambda$) | |
|---|---|---|
| | $1+\lambda+\lambda^4$ | $1+\lambda^3+\lambda^4$ |
| $1+\lambda^5+\lambda^6$ | 24 | 19 |
| $1+\lambda+\lambda^6$ | 6 | 18 |
| $1+\lambda^2+\lambda^3+\lambda^5+\lambda^6$ | 17 | 22 |
| $1+\lambda+\lambda^4+\lambda^5+\lambda^6$ | 21 | 16 |
| $1+\lambda+\lambda^3+\lambda^4+\lambda^6$ | 23 | 26 |
| $1+\lambda+\lambda^2+\lambda^5+\lambda^6$ | 19 | 28 |

## 4.0 CONCLUSIONS

It has been demonstrated through this simulation study that the change of the initial loading of PRTPG does not have any impact on the effectiveness of an LFSR based digital circuit testing technique that uses identical characteristic polynomials in both the PRTPG and MISR-SAZ as well. Thus, this result restricts the outright use of the results of findings; that the effectiveness of LFSR based digital circuit testing techniques can be improved by changing the initial loading of PRTPG. Thus, for effective use of initial loading of PRTPG of LFSR based digital circuit testing technique, it is essential to analyze the role of characteristic polynomials used in PRTPG as well as in MISR-SAZ. Although, our investigation is limited with small sizes of circuits but the trend of results suggests for further through analytical investigation.

**ACKNOWLEDGEMENTS**

The acknowledgements are due to authorities of Sultan Qaboos University (Sultanate of Oman) for providing generous research support grants and environments for carrying out the research works.

**Author** (Short Biography)

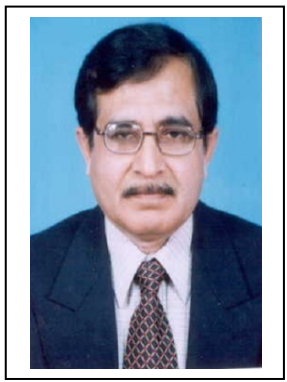

   **Afaq Ahmad** belongs to department of Electrical and Computer Engineering department at Sultan Qaboos University, Sultanate of Oman. He holds B.Sc. Eng., M.Sc. Eng., DLLR and Ph.D. degrees. Ahmad did his PhD from IIT Roorkee, India in 1990. Before joining Sultan Qaboos University, Dr. Ahmad was Associate Professor at Aligarh Muslim University, India. Prior to starting carrier at Aligarh, he also worked as consultant engineer with Light & Co., lecturer with REC Srinagar and senior research fellow with CSIR, India.

   **Dr. Ahmad** is Fellow member of IETE (India), senior member of IEEE Computer Society (USA) and life member of SSI (India), senior member IACSIT, member IAENG and WSEAS; He has published over 100 technical papers. At present he is associated as editors and reviewers of many reputed journals. He has delivered many keynote, invited addresses, extension lectures, organized conferences, short courses, and conducted tutorials at various universities of globally repute. He chaired many technical sessions of international conferences, workshops, symposiums, seminars, and short courses. He has undertaken and satisfactorily completed many highly reputed and challenging consultancy and project works. His research interests are: fault diagnosis and digital system testing, data security, graph theoretic approach, microprocessor based systems and computer programming.

   **Dr. Ahmad's** field of specialization is VLSI testing and fault-tolerant computing.